\documentclass{raa}

\usepackage{graphicx,times}             
\input{epsf.sty}                        
\input{psfig.sty}                       

\begin{document}

   \title{CCD Photometric Investigation of A W UMa-Type Binary GSC~0763-0572 
}

   \volnopage{Vol.0 (200x) No.0, 000--000}      
   \setcounter{page}{1}          

   \author{Wichean Kriwattanawong
      \inst{1,2}
   \and Porntipa Pooseekheaw
      \inst{1}
   }

   \institute{Department of Physics and Materials Science, Faculty of Science,
             Chiang Mai University, Chiangmai, 50200, Thailand; {\it k.wichean@gmail.com}\\
        \and
             National Astronomical Research Institute of Thailand, Chiangmai, 50200, Thailand\\
   }

   \date{Received~~2009 month day; accepted~~2009~~month day}

\abstract{ A photometric solution of an A-type W UMa binary, GSC~0763-0572 is examined with a revised orbital period. The overcontact degree is found to be $f$~=~40.66~$\%$, with a low mass ratio of ~$q$~=~0.2554. The result demonstrates an unambiguous increase in the orbital period with a relative period change of $\Delta P\slash P = +5.69\times10^{-7}~$d~yr$^{-1}$. This indicates that GSC~0763-0572 is undergoing a process of mass transfer from the secondary component to the primary one with a rate of relative mass change of $\Delta$$m_1$$\slash$$m$ $=+5.18\times10^{-8}$~yr$^{-1}$, for a conservative model of mass transfer. We find that GSC~0763-0572 might transform into a rapidly rotating star, if total spin angular momentum increases until it is greater than one-third of the orbital angular momentum, without breaking the contact configuration.
}

   \authorrunning{W. Kriwattanawong \& P. Pooseekheaw}            
   \titlerunning{CCD Photometric Investigation of A W UMa-Type Binary GSC 0763-0572
}  

   \maketitle

%
%
\section{Introduction}           
\label{sect:intro}

W UMa type systems are one of the most interesting types of binaries, which can be found in many studies that investigate their evolutionary states. These systems are believed to be formed from detached binaries, which are losing angular momentum and decreasing orbital period by a process called angular momentum loss (Vihu~\cite{VIHU81}; Rucinski \cite{RUCI82}; Guinan \& Bradstreet \cite{GUIN88}; Maceroni \& vant Veer \cite{MACE96}; Paczy\'nski et~al. \cite{PACZ06}). It is well-known that W~UMa binaries are classified into two subtypes (Binnendijk \cite{BINN70}). A W~UMa system is called A-subtype, if the more massive component is eclipsed by the less massive one at primary minima. On the other hand, a system with the opposite feature is called W-subtype. Some authors suggested that W-subtype systems have evolved from A-subtype ones, by mass loss (Gazeas \& Niarchos \cite{GAZE06}; Rucinski~\cite{RUCI85}), whereas others argued the opposite is true (Maceroni etal. \cite{MACE85}; Hilditch et al. \cite{HILD88}). However, van~Hamme~(\cite{VANH82}) and Li~et~al.~(\cite{LILI08}) found that there is no evolutionary difference between both subtypes.  Li et al. (\cite{LILI08}) suggested that the dynamical evolution would cause W UMa binaries to evolve into lower mass ratio systems and the tidal instability forces these systems to merge into rapidly rotating single stars.

An A-subtype W UMa binary, GSC 0763-0572 ($\alpha_{J2000}$=07$^{\rm h}$16$^{\rm m}$57$^{\rm s}$.32, $\delta_{J2000}$=09\dg 12'35''.5) was first discovered as a short-period variable by Bernhard (\cite{BERN02}).  Lloyd et al. (\cite{LLOY03}) reported five observations of light minima, displaying characteristics of a W UMa binary with an orbital period of 0.426388 d. Yang et al. (\cite{YANG05}) indicated that GSC 0763-0572 belongs to an A-subtype with a period of 0.4263947 d, slightly higher than that of Lloyd (\cite{LLOY03}). A photometric solution suggested that the system is a low mass ratio contact binary with $q$~=~0.2401, an orbital inclination of 79~\dg and a contact degree of 29.5~$\%$ (Yang et al. \cite{YANG05}).

This study presents a photometric study of GSC 0763-0572 with a revised orbital period and analysis of the period change and mass transfer rate. New observations and data reduction are first described in Section~2. The orbital period and the relative period change are examined in Section 3. Section 4 explains a light curve fit and a revised photometric solution of GSC 0763-0572. Finally, Sections 5 and 6 contain discussion and conclusions, respectively.

\section{Observations and Data reduction}
\label{sect:Obs}

Imaging data of GSC 0763-0572 were obtained from on the 0.5 m telescope at Sirindhorn Observatory, Chiang Mai University, Thailand. The observations were done for the $BVR$ filter bands, using an SBIG CCD (Model of ST10-XME), with exposure times of 60 seconds. The photometric data were taken during three nights (2010 December 29--31), covering approximately 2.5 periods. We have obtained 294 observations per filter band. The data were reduced, using the standard IRAF package.

$BVR$ differential magnitudes were determined using GSC 0763-0291 ($\alpha_{J2000}$=07$^{\rm h}$16$^{\rm m}$59$^{\rm s}$.49, $\delta_{J2000}$=09\dg 16'57''.6) and GSC 0763-0177 ($\alpha_{J2000}$=07$^{\rm h}$16$^{\rm m}$38$^{\rm s}$.98, $\delta_{J2000}$=09\dg 17'53''.0) as the comparison and check stars, respectively. The amplitude of light curve variation is about 0.45 mag, and the magnitude difference between the primary and the secondary eclipses is about 0.05 mag with almost the same level of light maxima.

\section{Orbital Period Change}
\label{sect:ROP}

The new photometric data of GSC 0763-0572 display four times of light minima in the $BVR$ filter bands. Light minimum times are determined in terms of HJD, using a least-squares method. New times of light minima are found as two primary and two secondary eclipses for each filter, listed in Table $\ref{min}$. The times of light minima in this study are combined with eight times from literature to estimate the orbital period, as shown in Table \ref{oc}. The HJD at light minimum times are fitted with the linear least-squares method and the result yields the orbital period of 0.4263965($\pm$0.0000002) d as shown in Equation~($\ref{min.eq}$). 

\begin{equation}
\rm Min.I = HJD 2455559.9439 (\pm 0.0009) + 0.4263965 (\pm 0.0000002) \times \it E 
\label{min.eq}
\end{equation}

\begin{table*}[h]
\begin{center}
\caption{\label{min} New times of light minima. }
\vspace{\baselineskip}
\begin{tabular}{lcccc} 
\hline
\hline
\multicolumn{1}{l}{No.} &
\multicolumn{1}{c}{Min } &
\multicolumn{1}{c}{Filter} &
\multicolumn{1}{c}{HJD} &
\multicolumn{1}{c}{Error} \\
\hline

1   &  II  &  B  &  2455560.1570  & 0.0010 \\
  &  II  &  V  &  2455560.1572  & 0.0009\\
  &  II  &  R  &  2455560.1572  & 0.0008\\
 2   &  ~I  &  B  &  2455560.3712  & 0.0016\\
  &  ~I  &  V  &  2455560.3700  & 0.0015\\
  &  ~I  &  R  &  2455560.3706  & 0.0007\\
 3   &  ~I  &  B  &  2455561.2234  & 0.0011\\
  &  ~I  &  V  &  2455561.2239  & 0.0021\\
  &  ~I  &  R  &  2455561.2234  & 0.0005\\
 4   &  II  &  B  &  2455561.4368  & 0.0014\\
  &  II  &  V  &  2455561.4369  & 0.0004\\
  &  II  &  R  &  2455561.4366  & 0.0007\\

\hline

\end{tabular}
\end{center}
\begin{tabular}{l}

\\

 Notes: Col. (1): the number of light minimum times. Col. (2): types of minima. Col. (3): filter bands.  Col. (4):\\
 HJD at light minima. Col. (5): errors of HJD at light minima.

\end{tabular}
\end{table*}

\begin{table*}[h]
\begin{center}
\caption{\label{oc} All light minimum times. }
\vspace{\baselineskip}
\begin{tabular}{lcccc} 
\hline
\hline

\multicolumn{1}{l}{HJD } &
\multicolumn{1}{c}{Epoch } &
\multicolumn{1}{c}{Min} &
\multicolumn{1}{c}{($O-C$)} &
\multicolumn{1}{c}{Ref.} \\
\hline
2452318.0524 & -7603.0 & ~I & 0.0015 & (1)\\
2452325.0886 & -7586.5 & II & 0.0021 & (1)\\
2452336.3876 & -7560.0 & ~I & 0.0016 & (1)\\
2452361.3321 & -7501.5 & II & 0.0019 & (1)\\
2452362.3974 & -7499.0 & ~I & 0.0012 & (1)\\
2453359.3092 & -5161.0 & ~I & -0.0019 & (2)\\
2453361.2282 & -5156.5 & II & -0.0017 & (2)\\
2453361.4387 & -5156.0 & ~I & -0.0044 & (2)\\
2455560.1571 & ~~~~0.5 & II & 0.0006 & (3)\\
2455560.3706 & ~~~~1.0 & ~I & 0.0009 & (3)\\
2455561.2235 & ~~~~3.0 & ~I & 0.0011 & (3)\\
2455561.4368 & ~~~~3.5 & II & 0.0012 & (3)\\
2455599.3872 & ~~~92.5 & II & 0.0023 & (4)\\

\hline

\end{tabular}
\begin{tabular}{l}
\\

 Notes: Col.(1): HJD at light minima. Col. (2): epoch. Col. (3): types of minima. Col. (4): residuals of \\
 HJD at light minima. Col. (5): references for sources are as follows: [1] Lloyd et al. (\cite{LLOY03}); [2] Yang\\  et al. (\cite{YANG05}); [3] This study; [4] H\"ubscher (\cite{HUBS11}).
\end{tabular}
\end{center}
\end{table*}

The orbital period of GSC 0763-0572 shows progressive lengthening from the values obtained by Lloyd et al. (\cite{LLOY03}), Yang~et~al.~(\cite{YANG05}) and our study. With this orbital period increase, the residuals ($O-C$) are determined as listed in Table $\ref{oc}$ and the least-squares fitting solution yields Equation ($\ref{o-c.eq}$).  It is found that the orbital period increases with a relative change of $\Delta P\slash P = +5.69\times10^{-7}~$d~yr$^{-1}$. However, the number of light minimum times is not very large. Thus, more observations are needed in the future.

\begin{equation}
(O-C) = 0.00032 (\pm 0.00038) + 2.38 (\pm 0.31) \times 10^{-6} E + 3.32(\pm 0.41) \times 10^{-10} E^2 
\label{o-c.eq}
\end{equation}

\section{Light Curve fit}
\label{LCF}

Observed light curves in the $BVR$ filter bands are fitted with the 2003 version of the Wilson-Devinney (W-D) code (Wilson \& Devinney \cite{WILS71}; Wilson \cite{WILS79}, \cite{WILS90}; Wilson \& van Hamme \cite{WILS03}). The applied orbital period is 0.4263965 d. Total magnitudes of GSC 0763-0572 in the $B$ and $V$ filter bands from the Tycho-2 Catalog (Hog et al. \cite{HOGE00}) are 11.423 and 10.673, respectively. Thus the color-temperature relation (Sekiguchi and Fugita \cite{SEKI00}; Ram\'irez and Mel\'endez \cite{RAMI05}) gives the estimated temperature of the primary star of approximately 5,300 K.  Required parameters are adopted for the W-D code in mode 3, as listed in Table \ref{param}. The gravity darkening exponents of both components are equal, $g_1~=~g_2$~=~0.32 (Lucy \cite{LUCY67}), and the bolometric albedo coefficients of star 1 and star 2 are given as 0.50, $A_1~=~A_2$=~0.50 (Rucinski \cite{RUCI73}).  The limb darkening coefficients of star1 and star 2 are adopted as $x_{1B}$~=~0.840, $x_{2B}$~=~0.830, $x_{1V}~=~x_{2V}$~=~0.750 and $x_{1R}~=~x_{2R}$~=~0.670 (Al-Naimiy \cite{ALNA78}). The adopted adjustable parameters are the orbital inclination, $i$, estimated secondary temperature, $T_2$, mass ratio, $q$, the surface potential of the components, $\Omega_1~=~\Omega_2$, and the monochromatic luminosity of star 1, $L_1$. The relative luminosity of star 2 is estimated, using a model of stellar atmospheres (Kurucz \cite{KURU93}).

\begin{table*}[h]
\begin{center}
\caption{\label{param} Photometric solution for the light curve fit of GSC 0763-0572.}
\vspace{\baselineskip}
\begin{tabular}{lr} 
\hline
\hline

\multicolumn{1}{l}{Parameters } &
\multicolumn{1}{r}{Values} \\
\hline

$i$($^{\circ}$)  & 79.10(0.43) \\
$g_1~=~g_2$  & 0.32   \\
$A_1~=~A_2$ & 0.50 \\
$T_1$($K$) & 5300\\
$T_2$($K$) & 5256(30) \\
$\Omega_1~=~\Omega_2$ & 2.3000(0.0051) \\
$q$ & 0.2554(0.0035) \\
$L_{1B}/( L_{1B}~+~L_{2B} )$ & 0.7755(0.0023) \\ 
$L_{1V}/( L_{1V}~+~L_{2V} )$ & 0.7779(0.0022) \\ 
$L_{1R}/( L_{1R}~+~L_{2R} )$ & 0.7769(0.0022) \\
$r_1$ (pole)  &  0.4831(0.0014) \\
$r_1$ (side)  &  0.5263(0.0020) \\
$r_1$ (back)  &  0.5555(0.0029) \\
$r_2$ (pole)  &  0.2657(0.0037) \\
$r_2$ (side)  &  0.2789(0.0046) \\
$r_2$ (back)  &  0.3276(0.0106) \\
$x_{1B} $ & 0.840\\
$x_{2B}$ & 0.830\\
$x_{1V}$  & 0.740\\
$x_{2V}$  & 0.740\\
$x_{1R}$  & 0.670\\
$x_{2R}$  & 0.670\\
$\Sigma$($O-C$)$^2$   & 0.0040\\
$f~(\%$)&     40.66(3.14) \\

\hline

\end{tabular}
\end{center}
\end{table*}

\begin{figure}[t]
\centering
\vspace*{.6cm}
\includegraphics[width=0.60\textwidth,height=0.50\textheight,angle=-90]{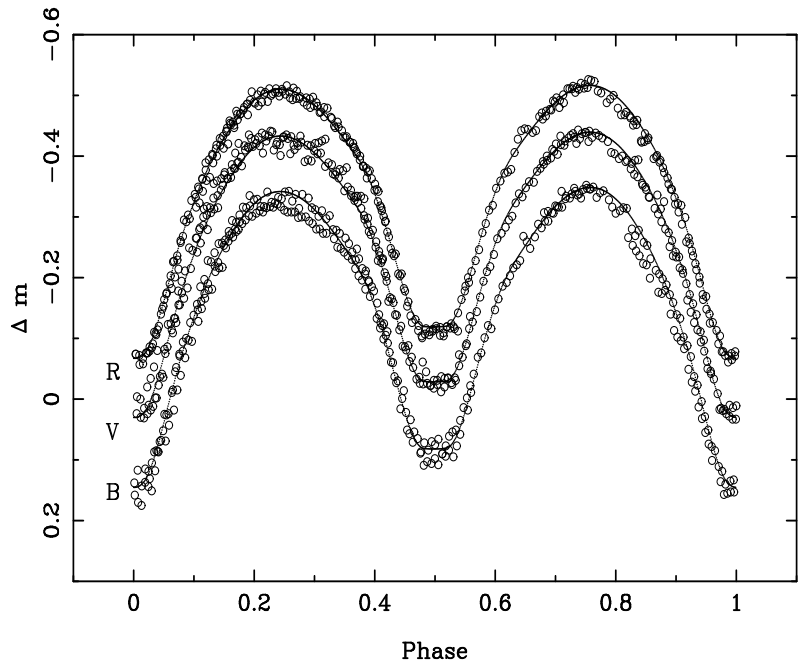}
\caption{Observed in the $B$ (circle), $V$ (square) and $R$ (triangle) filter bands and theoretical (solid lines) light curves vs. orbital phase.}
\label{lc}
\end{figure}

A good fit with the W-D code gives a mass ratio of $q$~=~0.2554($\pm$0.0035), and the overcontact degree of $f$~=~40.66$\%$($\pm$3.14$\%$), which is higher than the previous study of $f$~=~29.5~$\%$ (Yang et al. \cite{YANG05}). Meanwhile, radius parameters of the secondary component, $r_2$ (pole) = 0.2657(0.0037), $r_2$ (side) =~0.2789(0.0046) and $r_2$ (back) = 0.3276(0.0106), slightly increase and are comparable to the values of Yang et al. (\cite{YANG05}), while the radius parameters of the primary one do not significantly change. The temperature difference between the components is not very large (i.e., $T_1$~=~5300~K and $T_2$~=~5256~K). The sum of squares of residuals for input values, $\Sigma$($O-C$)$^2$~=~0.0040, for corresponding to the light curves, and all main parameters are listed in Table $\ref{param}$. Finally, the theoretical light curves are plotted, overlaid on the observed ones, as shown in Figure $\ref{lc}$.

\section{Discussion}
\label{sect:discussion}

The photometric solution from W-D synthesis with $q$~=~0.2554($\pm$0.0035) and the theoretical light curves in Figure $\ref{lc}$ clearly show that GSC 0763-0572 is an A-subtype W-UMa binary, as Yang et al. (\cite{YANG05}) mentioned. The orbital period of GSC 0763-0572 is gradually increasing from 0.426388 d (Lloyd et al. \cite{LLOY03}) and 0.4263947($\pm$0.0000003) d (Yang et al. \cite{YANG05}) to 0.4263965($\pm$0.0000002) d in this study.

The orbital period of GSC 0763-0572 increases at a rate of the relative period change, $\Delta P$$\slash$$P$ $= +5.69$$\times$$10^{-7}$~d~yr$^{-1}$. This indicates that the system is undergoing mass transfer from the less massive secondary to the more massive primary component with a relative mass change, $\Delta$$m_1$$\slash$$m$ $= +5.18\times10$~$^{-8}$~yr$^{-1}$, for a conservative model of mass transfer (Pribulla \cite{PRIB98}; Pribulla et al. \cite{PRIB99}; Singh and Chaubey \cite{SING86}), where $\Delta {m_1}$ is the mass increase of the more massive component ($\Delta {m_1} = -\Delta {m_2}$), and $m$ is total mass ($m = m_1 + m_2$). This process of mass transfer is similar to the results, found in V700 Cyg (Yang et al. \cite{YANG09b}) and V345~Gem (Yang et al. \cite{YANG09c}), and also in the Algol-type VW Hya (Zhang~et~al. \cite{ZHAN09}), and the $\beta$-Lyr type AI~Cru (Zhao etal. \cite{ZHAO10}).

As the orbital period increase, the corresponding separation between both components spreads out. This process is caused by orbital angular momentum loss and the mass transfer enforces the system to have smaller mass ratio, thus increasing the spin angular momentum. It is possible that GSC 0763-0572 could satisfy the condition that the total spin angular momentum exceed one-third of the orbital angular momentum, i.e., 3$J_{spin} > J_{orb}$, so the system will be unstable (Hut \cite{HUTP80}). The resulting tidal instability compels the system to evolve into a rapidly rotating single star (Li et al. \cite{LILI08}), similar to what was found in the analysis of QX And (Qian et al \cite{QIAN07}), V343 Ori (Yang et al. \cite{YANG09a}) and V345 Gem (Yang et al. \cite{YANG09c}).

\section{Conclusions}
\label{sect:conclusion}

New photometric data of GSC 0763-0572 show four times of light minima. Combining the light minimum times from literature, and using the least squares method, we give a revised orbital period of 0.4263965($\pm$0.0000002) d. Although the number of light minimum times is not very large, the result clearly exhibits an increase in the orbital period. The photometric solution of the A-subtype W~UMa binary GSC 0763-0572 is deduced, providing some revised parameters (i.e. the mass ratio, $q$~=~0.2554 and the overcontact degree, $f$~=~40.66~$\%$), which are higher than those in the literature. 

The relative change of the orbital period of GSC 0763-0572 is found to be the rate of $\Delta P\slash P$ $=~+5.69~\times~10^{-7}~$d~yr$^{-1}$. The period increase is being caused by mass transfer from the less massive component to the more massive one with a relative mass change of $\Delta {m_1}\slash m = +5.18~\times~10^{-8}~$yr$^{-1}$, for a conservative model. Because component separation increases while mass ratio decreases during expansion of the secondary, GSC 0763-0572 could evolve into a rapidly-rotating star, when the system reaches the condition of Hut (\cite{HUTP80}). However, there are not many recorded times of light minima for this system, so more observations are still needed in the future to examine the long-term period change, which can more precisely, confirm how GSC~0763-0572 evolves.

\begin{acknowledgements}

We acknowledge partial support by the Thailand International Development Cooperation Agency (TICA). New observations of GSC 0763-0572 were obtained using the 0.5 m telescope at Sirindhorn Observatory, Chiang Mai University. This research has made use of the SIMBAD online database, operated at CDS, Strasbourg, France and the Astrophysics Data System (ADS), operated by the Smithsonian Astrophysical Observatory (SAO) under a NASA grant.

\end{acknowledgements}

\label{lastpage}

\end{document}